\documentclass[11pt]{article}
\usepackage[margin=1.0in]{geometry}
\usepackage{graphicx,picinpar}               
\usepackage[pdftex,hyperfootnotes=false,plainpages=false,pdfpagelabels=true]{hyperref}
\usepackage{latexsym}
\usepackage{alltt} 
\usepackage{verbatim} 
\usepackage{amsmath}
\usepackage{times}
\usepackage{amsthm}
\usepackage{amsfonts}
\usepackage{amssymb}
\usepackage{upgreek}
\usepackage{floatflt}
\usepackage{wrapfig}
\usepackage{multicol}
\usepackage{subfigure}
\usepackage{xspace}
\usepackage[numbers]{natbib}
\usepackage{paralist}
\usepackage{color}

\begin{document}

\title{\bf The Whole is Greater than the Sum of the Parts: \\
Optimizing the Joint Science Return from LSST, Euclid and WFIRST}
\maketitle

\begin{quote}
{B.~Jain,\footnote{bjain@physics.upenn.edu} D.~Spergel,\footnote{dns@astro.princeton.edu} 
R.~Bean, A.~Connolly, I.~Dell'antonio, J.~Frieman, E.~Gawiser, N.~Gehrels, L.~Gladney, K.~Heitmann, 
G.~Helou, C.~Hirata, S.~Ho, \v{Z}.~Ivezi\'{c}, M.~Jarvis, S.~Kahn, J.~Kalirai, A. Kim, R. Lupton, 
R.~Mandelbaum, P.~Marshall, J.~A.~Newman, S.~Perlmutter, M.~Postman, J.~Rhodes, M.~A.~Strauss, 
J.~A.~Tyson, L.~Walkowicz, W.~M.~Wood-Vasey}
\end{quote}

\tableofcontents

\section{Where will we be in 2024?}
Astronomy in 2024 should be very exciting!  LSST and Euclid, which should each be in the midst of their deep surveys of the sky,
will be joined by WFIRST. With higher resolution and sensitivities than previous astronomical survey instruments, they will reveal new insights into areas ranging from exoplanets to the nature of dark energy.  At the same time, JWST will be staring deeper into the early universe than ever before.  Advanced LIGO should be detecting frequent collisions between neutron stars.  ALMA will be operating at all of its planned frequencies, and the new generation of very large optical ground based telescopes should be revolutionizing ground-based optical astronomy.   In parallel, advances in computational capabilities should enable observers to better exploit these complex data sets and theorists to make detailed time-dependent three-dimensional models that can capture much of the physics needed to explain the new observations.

The focus of this report is an exploration of some of the opportunities enabled by the combination of LSST, Euclid and WFIRST, the optical surveys that will be an essential part of the next decade's astronomy.  The sum of these surveys has the potential to be significantly greater than the contributions of the individual parts.  As is detailed in this report, the combination of these surveys should give us multi-wavelength high-resolution images of galaxies and broadband data covering much of the stellar energy spectrum.  These stellar and galactic data have the potential of yielding new insights into topics ranging from the formation history of the Milky Way to the mass of the neutrino.  However, enabling the astronomy community to fully exploit this multi-instrument data set is a challenging technical task: for much of the science, we will need to combine the photometry across multiple wavelengths with varying spectral and spatial resolution. Coordination will be needed between the LSST, Euclid, and WFIRST projects in order to understand the trades between overlapping areal coverage, filter design, depth and cadence of the observations, and performance of the image analysis algorithms.
We will need to provide these data to the community in a highly usable format.  If we do not prepare the missions for this task in advance, we will limit their scientific return and increase the cost of the eventual effort of fully exploiting these data sets.  The goal of this report is to identify some of the science enabled by the combined surveys and the key technical challenges in achieving the synergies.

\newpage
\section{Background on missions}
This section covers the science goals and technical overview of the three missions.

\subsection{WFIRST}

WFIRST is the top-ranked large space mission of the 2010 New Worlds New Horizon (NWNH) Decadal Survey.  With the addition of a coronagraph, WFIRST would also satisfy the top ranked medium space priority of NWNH. The mission is designed to settle essential questions in dark energy, exoplanets, and infrared astrophysics.  The mission will feature strategic key science programs plus a vigorous program of guest observations.  A Science Definition Team and a Study Office at the Goddard Space Flight Center and Jet Propulsion Laboratory are studying the mission with reports\footnote{See SDT web site:  {\it http://WFIRST.gsfc.nasa.gov/add/} } in 2013, 2014, and 2015.

WFIRST is now baselined with an existing 2.4 m telescope NASA acquired from the National Reconnaissance Office (NRO), a telescope that became available after the completion of NWNH.  This configuration is referred to as WFIRST-AFTA (Astrophysics Focused Telescope Asset).  The mission consists of telescope, spacecraft, a Wide-Field Instrument (WFI), an IFU spectrometer and a coronagraph instrument, which includes an IFS spectrometer.  The WFI operates with multiple bands covering  0.7 to 2.0 micron band with extension to 2.4 microns under study.  It has a filter wheel for multiband imaging and a grism for wide-field spectroscopy (R=550-800).  The pixel scale is 0.11 arcsec, which fully samples the PSF at the H band. The IFU has a 3 arcsec FoV and R=75 resolution.  The coronagraph operates in the visible 400 - 1000 nm band.  It has a 2.5 arcsec FoV, $10^{-9}$ effective contrast and 100-200 mas inner working angle.  The IFS has R=70 resolution.

WFIRST will measure the expansion history of the Universe as a function of cosmic time and the growth rate of the large-scale structure of the Universe  as a function of time to test the theory of general relativity on cosmological scales and to probe the nature of dark energy.  It will employ five different techniques: supernovae, weak lensing, baryon acoustic oscillations (BAO), redshift space distortions (RSD), and clusters.  With its 2.4-meter primary mirror, the mission will measure more than double the surface density of galaxies detected by the Euclid mission and push these measurements further in the NIR.  This higher density will enable more detailed maps of the dark matter distribution, measurements not only of two-point statistics but also higher order statistics, and multiple tests of systematic dependance of cosmological parameters on galaxy properties. The NRC noted in its recent review\footnote{{\it http://www.nap.edu/catalog.php?record\_id=18712}}: ``For each of the cosmological probes described in NWNH, WFIRST/AFTA exceeds the goals set out in NWNH."

For exoplanets, WFIRST will complete the census of exoplanets begun by Kepler.  It will make microlensing observations of the galactic bulge using the WFI to discover several thousand planets at the orbit of Earth and beyond and will be sensitive to planets as small as Mars. WFIRST's microlensing survey will have the unique capability to detect free-floating planets, thus, enabling astronomers to determine the efficiency of planet formation.  The coronagraph will directly image and characterize tens of Uranus to Jupiter mass planets around nearby stars and study debris disks.

WFIRST will also be a worthy successor to the Hubble Space Telescope.  With 200 times the field-of-view of HST, and the same size primary mirror, it will conduct a rich program of general astrophysics and reveal new insights and discoveries on scales ranging from the nearest stars to the most distant galaxies.

Table 1 gives the WFIRST capabilities for surveys and SN monitoring observations.

\begin{table}[h!]
\centering
\begin{tabular}[b]{| l | l |}
\hline
\textbf{Attributes}					&	\textbf{WFIRST Capability}					\\ \hline
Imaging survey						&	J $\sim$ 27 AB  over 2400 sq deg			\\ \hline
									&	J $\sim$ 29 AB over 3 sq deg deep fields	\\ \hline
Multi-filter photometry				&	Filters:  z, Y, J, H, F184 (Ks), W (wide)	\\ \hline
Slitless wide-field spectroscopy	&	0.28 sq deg,  R$\sim$600					\\ \hline
Slit multi-field spectroscopy		& 	IFU,  R$\sim$70								\\ \hline
Number of SN Ia 					& 	$2\times10^{3}$ to z$\sim$1.7				\\ \hline
Number galaxies with spectra		& 	$2\times10^{7}$								\\ \hline
Number galaxies with shapes			& 	$4\times10^{8}$								\\ \hline
Number of galaxies detected			& 	few $\times10^{9}$							\\ \hline
Number of massive clusters			& 	$4\times10^{4}$								\\
\hline
\end{tabular}
\caption{\textbf{WFIRST capabilities in a nominal $\sim2.5$ year dark energy survey}}
\label{table:1}
\end{table}

{\it High Latitude Survey (HLS):}
The nominal HLS will be for 2 years with 1.3 years for imaging and 0.6 years for grism spectroscopy.  The coverage will be 2200 deg$^{2}$ of high Galactic latitude sky within the LSST footprint.  The imaging will have 2 passes over the survey footprint in each of the 4 imaging filters (J, H, F184 [for shapes] and Y [for photo-z's]).  Data from LSST will be required to provide optical filters for photo-z determination.  Each pass will include four 184 sec exposures in each filter (with five exposures in J band) with each exposure offset diagonally by slightly more than a detector chip gap. This pattern is repeated across the sky.
The spectroscopic survey will have four passes total over the survey footprint with two ``leading" passes and two ``trailing" passes to enable the single grism to rotate relative to the sky.  Each pass includes two 362 sec exposures with offset to cover chip gaps.  The two ``leading" passes (and two ``trailing") are rotated from each other by $\sim5^{\circ}$.  
It is anticipated that the sky coverage of the HLS will be significantly expanded in the extended phase of the mission after the baseline first 6 years. WFIRST has no consumables that would prohibit a mission of 10 years or longer, and is being designed with serviceability in mind to enable the possibility of an even longer mission.

{\it Supernova Survey:}
A six month SN Ia survey will employ a three-tier strategy so to track supernova over a wide range of redshifts:
\begin{itemize}
\item Tier 1 for z$<$0.4:   27.44 deg$^{2}$  Y=27.1, J= 27.5
\item Tier 2 for z$<$0.8: 8.96 deg$^{2}$ J=27.6, H= 28.1
\item Tier 3 for z$<$1.7: 5.04 deg$^{2}$ J=29.3, H=29.4
\end{itemize}
Tier 3 is contained in Tier 2 and Tier 2 is contained in Tier 1.  Each of these fields will be visited every 5 days over the 6 months of the SN survey.  The imager is used for SN discovery and the IFU spectrometer is used to determine SN type, measure redshifts, and obtain lightcurves.  Each set of observations will take a total of 30 hours of combined imaging and spectroscopy.  The fields are located in low dust regions $\le20^{\circ}$ off an ecliptic pole.  A final revisit for each target for spectroscopy will occur after the SN fades for galaxy subtraction.

{\it Guest Observer Program:}
A significant amount of observing time will be awarded to the community through a peer-selected GO program.  An example observing program prepared by the SDT has 25\% of the baseline 6 years of the mission, or 1.5 years, for guest observations.  Significantly more time would be awarded in an extended phase after the 6th year.  The GO program is expected to cover broad areas of science from the solar system to Galactic studies to galaxies to cosmology.  The April 2013 SDT study  contains a rich set of $\sim$50 potential GO science programs that are uniquely enabled by WFIRST.  Astronomical community members provided a wide-ranging set of one-page descriptions of different GO programs that highlight the tremendous potential of WFIRST to advance many of the key science questions formulated by the Decadal survey.  An important contribution that would likely come from key projects in the GO program would be to have deep IR observations of selected few-square-degree fields.  These ultra deep drilling fields could reach limits of J=30 AB.

\subsection{LSST}
The Large Synoptic Survey Telescope (LSST) is a large-aperture, wide-field, ground-based facility designed to perform many repeated imaging observations of the entire southern hemisphere in six optical bands (u, g, r, i, z, y).  The majority of the southern sky will be visited roughly 800 times over the ten-year duration of the mission.  The resulting database will enable a wide array of diverse scientific investigations ranging from studies of moving objects in the solar system to the structure and evolution of the Universe as a whole (Ivezic et al 2014). 

The Observatory will be sited atop Cerro Pachon in Northern Chile, near the Gemini South and SOAR telescopes.   The telescope incorporates a 3-mirror astigmatic optical design. Incident light is collected by the primary, which is an annulus with an outer diameter of 8.4 m, then reflected to a 3.4-m convex secondary, onto a 5-m concave tertiary, and finally into three refractive lenses in the camera. The total field of view is 9.6 deg$^2$ and the effective collecting aperture is 6.6 m in diameter. The design maintains a 0.2 arc-second system point spread function (PSF) across the entire spectral range of 320 nm to 1050 nm.  The etendue (the product of collecting area and field of view) of the system is several times higher than that of any other previous facility.

The telescope mount assembly is a compact, stiff structure with a high fundamental frequency that enables fast slew and settle.   The camera contains a 3.2 Gigapixel focal plane array, comprised of roughly 200 4K $\times$ 4K CCD sensors with 10 $\mu$m pixels. The sensors are deep depleted, back-illuminated devices with a highly segmented architecture that enables the entire array to be read out in 2 s or less.

Four major science themes have motivated the definition of science requirements for LSST:

\begin{itemize}
\item	Taking a census of moving objects in the solar system.

\item	Mapping the structure and evolution of the Milky Way.

\item	Exploring the transient optical sky.

\item	Determining the nature of dark energy and dark matter.
\end{itemize}

These four themes stress the system in different ways.  For the science of dark energy and dark matter, LSST data will enable a variety of complementary analyses, including measurements of cosmic shear power spectra, baryon acoustic oscillations, precision photometry of Type Ia supernovae, measurements of time-delays between the multiple images in strong lensing systems, and the statistics of clusters of galaxies.  Collectively, these will result in substantial improvements in our constraints on the dark energy equation of state and the growth of structure in the Universe, among other parameters.

The main fast-wide-deep survey will require 90\% of the observing time and is designed to optimize the homogeneity of depth and number of visits. Each visit will comprise two back-to-back 15 second exposures and, as often as possible, each field will be observed twice per night, with visits separated by 15-60 minutes. Additional survey areas, including a region within 10 degrees of the northern ecliptic, the South Celestial Pole, and the Galactic Center will be surveyed with either a subset of the LSST filter complement or with fewer observations. The remaining 10\% of the observing time will be used on mini-surveys that improve the scientific reach of the LSST. Examples of this include the ``Deep Drilling Fields'' where each field receives approximately 40 hour-long sequences of 200 exposures. When all 40 sequences and the main survey visits are coadded, this would extend the depth of these fields to $r\sim28$. The LSST has identified four distant extragalactic survey fields to observe as Deep Drilling Fields: Elias S1, XMM-LSS, Extended Chandra Deep Field-South, and COSMOS.

Table~\ref{table:2} gives the LSST baseline design and survey parameters  

\begin{table}[h]
\centering
\begin{tabular}{|l|l|}
\hline  
\textbf{Attributes}					&	\textbf{LSST Capability}					\\ 
\hline  
Final f-ratio, aperture                 &  f/1.234, 8.4 m                \\ 
Field of view, \'etendue              &  9.6 deg$^2$,   319 m$^2$deg$^2$     \\ 
Exposure time                            & 15 seconds (two exposures per visit)\\  
Main survey area                        & 18,000 deg$^2$\\  
Pixel count                                  &  3.2 Gigapix  \\ 
Plate scale                                  &  50.9 $\mu$m/arcsec (0.2'' pix)  \\  
Wavelength coverage                   &  320 -- 1050 nm, $ugrizy$             \\ 
Single visit depths, design   &  23.7, 24.9, 24.4, 24.0, 23.5, 22.6    \\ 
Mean number of visits          &  56, 80, 184, 184, 160, 160               \\  
Final (coadded) depths         &  25.9, 27.3, 27.2, 26.8, 26.3, 25.4     \\
\hline  
\end{tabular}
\caption{\textbf{LSST  Capabilities for the fast-wide-deep main survey}}
\label{table:2}
\end{table}

The rapid cadence of the LSST will produce an enormous volume of data, roughly 15 Terabytes per night, leading to a total dataset over the ten years of operations of over a hundred Petabytes. Processing such a large dataset and archiving it in useful form for access by the community has been a major design consideration for the project. The data management system is configured in three layers: an infrastructure layer consisting of the computing, storage, and networking hardware and system software; a middleware layer, which handles distributed processing, data access, user interface, and system operations services; and an applications layer, which includes the data pipelines and products and the science data archives. There will be both a mountain summit and a base computing facility (located in La Serena, the city nearest the telescope site), as well as a central archive facility in the United States. The data will be transported over existing high-speed optical fiber links from South America to the U.S.

The observing strategy for the LSST will be optimized to maximize observing efficiency by minimizing slew and other down time and by making appropriate choices of filter bands given the real-time weather conditions. The final cadence selection will be undertaken in consultation with the community.  A prototype simulator has been developed to help evaluate this process, which will be transformed into a sophisticated observation scheduler. A prototype fast Monte Carlo optical ray trace code has also been developed to simulate real LSST images. This will be further developed to aid in testing science analysis codes. The LSST cadence will take into account the various science goals and can also be refined to improve the synergy with other datasets.

LSST anticipates first light in 2020 and the start of the 10-year survey in 2022. It is fully funded for construction as of the summer 2014.

\subsection{Euclid}

Euclid is a medium class mission within the European Space Agency's (ESA) `Cosmic Vision' program\footnote{ {\it http://sci.esa.int/cosmic-vision/}}.  Euclid was formally selected in 2011 and is currently on schedule for launch in 2020. After traveling to the L2 Lagrange point and a brief shakeout and calibration period, Euclid will undertake an approximately 6-year survey aimed at ``mapping the geometry of the dark Universe." Euclid will pursue four primary science objectives (Laureijs et al 2011):
\begin{itemize}
\item 	Reach a dark energy FoM $> 400$ using only weak lensing and galaxy clustering; this roughly corresponds to 1\% errors on $w_p$ and $w_a$ of 0.02 and 0.1, respectively.
\item 	Measure $\gamma$, the exponent of the growth factor, with a 1$\sigma$ precision of $< 0.02$, sufficient to distinguish General Relativity and a wide range of modified-gravity theories.
\item 	Test the Cold Dark Matter paradigm for hierarchical structure formation, and measure the sum of the neutrino masses with a 1$\sigma$ precision better than 0.03 eV.
\item 	Constrain $n_s$, the spectral index of the primordial power spectrum, to percent accuracy when combined with Planck, and to probe inflation models by measuring the non-Gaussianity of initial conditions parameterized by
$f_{NL}$ to a 1$\sigma$ precision of $\sim$2.
\end{itemize}

Euclid is optimized for two primary cosmological probes:  weak gravitational lensing and galaxy clustering (including BAO and RSD).  Therefore, Euclid will measure both the expansion history of the Universe and the growth of structure. Euclid comprises a 1.2m Korsch 3 mirror anastigmatic telescope and two primary science instruments.  The visible instrument (VIS) consists of 36 4k $\times$ 4k CCDs and will be used to take images in a single, wide (riz) filter for high precision weak lensing galaxy shape measurements.  The light entering the telescope will be split via a dichroic to allow for simultaneous observations with the Near Infrared Spectrometer and Photometer (NISP), which will take 3 band imaging (Y, J, H) and grism spectroscopy in the 1-2 $\mu$m wavelength range. The NISP imaging is aimed at producing high quality photometric redshifts when combined with ground-based optical data from a combination of telescopes in the southern and norther hemispheres and the spectroscopy is aimed at producing accurate maps of galaxy clustering over 2/3 of the age of the Universe. NISP will contain 16 H2RG 2k $\times$ 2k NIR detectors procured and characterized by NASA.  Together, VIS and NISP will survey the darkest (least obscured by dust) 15,000 square degrees of the extragalactic sky, providing weak lensing shapes for over 1.5 billion galaxies and emission line spectra for several tens of millions of galaxies, while taking full advantage of the low systematics afforded by a space mission.

While the Euclid hardware and survey design are specifically optimized for dark energy studies, the images and catalogs will enable a wide range of ancillary science in cosmology, galaxy evolution, and other areas of astronomy and astrophysics.  These data will be made publicly available both in Europe via ESA and The Euclid Consortium and in the US via the Euclid NASA Science Center at IPAC (ENSCI) after a brief proprietary period.  Small areas of the Euclid survey (suitable for testing algorithms and pipelines but not large enough for cosmology) will be released at 14, 38, 62, and 74 months after survey operations start (these are referred to as ``quick releases").  The full survey data will be released in three stages:  circa 2022, 2500 square degrees; circa 2025, 7500 square degrees; circa 2028, 15,000 square degrees.

\newpage
\section{Enhanced science}

The main dark energy probes established over the last decade are described next in the context of the three missions. They are: Type Ia supernova, galaxy clustering (baryon acoustic oscillations and redshift space distortions), weak lensing, strong lensing and galaxy clusters. The biggest advantage from combining information from the three missions is in the mitigation of systematic errors, especially via redshift information.  For the observational and theoretical issues underlying the discussion here we refer the reader to recent review articles in the literature (Weinberg et al 2013; Joyce et al 2014). 

LSST, WFIRST and Euclid each have their own systematic errors, most of which are significantly reduced via the combination of the  survey datasets. The systematic errors affecting each of the surveys individually  arise  from their incomplete wavelength coverage (creating photo-z systematics), their differences in imaging resolution and blending (creating shear systematics), or from different biases in galaxy sample selections.   In most respects the  surveys are complementary in the sense that one survey reduces or nearly eliminates a systematic in the other.
To achieve this synergy requires some level of data sharing; we discuss below the cases where catalog level sharing of data is sufficient and others where it is essential to jointly process the pixel data from the surveys.

This section describes each dark energy probe, with a focus on mitigation of systematics, and returns to photometric redshifts (photo-z's). The last three subsections are devoted to stellar science, galaxy and quasar science, and the variable universe.

\subsection{Weak lensing}

A cosmological weak lensing analysis in any of these surveys will use tomography, which involves
dividing the galaxy sample into redshift slices and measuring both the auto-correlations of
galaxies within slices and cross-correlations of galaxies in different slices.  Tomography gives
additional information about structure growth beyond a strictly 2D shear analysis.  However,
there are a number of requirements on the data for a tomographic weak lensing analysis to be
successful.  First, we require a good understanding of both additive and multiplicative bias in the shear estimates, including how they scale with redshift.  Second, we require a strong
understanding of photometric redshift bias and scatter.  This is true in general, but it becomes
even more important when trying to model theoretical uncertainties such as the intrinsic
alignments of galaxy shapes with large-scale structure.

There are a number of ways in which the weak lensing science from each of LSST, WFIRST and
Euclid would benefit by sharing data with the other surveys and thus improving our understanding
of shear estimates and/or the photometric redshift estimates.  This data sharing may include
catalog-level data or possibly image data.  Although the latter would pose additional
complications in terms of the data reprocessing required to appropriately incorporate both
ground and space images, the gains may be worth the effort where confusion or depth or wavelength limitations are severe.


The main benefit to LSST from either WFIRST or Euclid will be in terms of the calibration of
shear estimates.  The space-based imaging will have much better resolution due to the much
smaller PSF size.  The shapes of galaxies measured from space images will suffer less from model
bias and noise bias (at a fixed depth) than the shapes of the same galaxies measured from ground
images.  LSST  does go deeper over its wide survey than the space missions, so it will do better on the outer parts of galaxy images. For subsets of galaxies measured at comparable depth, a comparison of the shear signal estimated by both LSST and either Euclid or WFIRST can  help quantify  possible systematic errors in the LSST
estimates.  While each survey will have its own independent scheme to correct for calibration
biases, a cross-comparison of this sort can be used to validate those schemes.  Any additional
correction that is derived could then be extended to the rest of the LSST area where there is no
such overlap.  Note that such a comparison would use a matched sample of galaxies in LSST and
one of the space-based surveys, but would not involve comparing per-galaxy shape estimates.
With different effective resolution, wavelength, and possibly shear estimation methods, there is
no reason to expect agreement in per galaxy shears (even allowing for variation due to noise).
Instead, such a comparison would utilize the reconstructed lensing shear fields from the
different surveys.  WFIRST may be more effective than Euclid for this purpose because its
narrower passbands mean that it does not suffer from the wide-band chromatic PSF issue mentioned
below. 

Another benefit to LSST relates to blended galaxies.  The higher resolution
space-based images make it easier to reliably identify blended galaxies than in ground-based images.
The performance of the deblending algorithm in LSST image processing pipeline can be significantly 
improved by providing higher-resolution space-based data because of order half of all observed galaxies
(though a smaller fraction of the LSST gold sample) will be significantly blended with another galaxy.  With space-based catalogs, one can
at least identify which objects are really multiple galaxies or galaxy-star blends and do forced fitting
using the space-estimated positions.  Euclid will be more effective than WFIRST for this
purpose because its footprint has a greater overlap with that of LSST.

One benefit to Euclid from combination with LSST relates to chromatic effects. The
diffraction-limited PSF is wavelength dependent and therefore differs for stars and galaxies due
to their different SEDs.  Euclid's very wide band imaging means that with its optical imaging
alone, it will not be able to correctly estimate the appropriate PSF for each galaxy.
Multi-band photometry is required to correct for this effect.  Without correctly accounting for
the color-dependent PSF, the systematic will dominate Euclid's WL error budget, so LSST photometry
will be very useful in reducing this systematic error.  Note that LSST cannot help with chromatic effects
involving color gradients within galaxies, for which higher resolution imaging data will be necessary to estimate
corrections; but LSST can provide information about chromatic effects involving the average
galaxy colors.

Finally, there are some shear systematics tests that can be done in a joint analysis but not in
any one survey individually.  For example, one route that has been proposed within surveys to
mitigate additive shear systematics is to cross-correlate shear estimates in different bands or
in different exposures rather than auto-correlate shear estimates.  However, some sources of
additive systematics may persist across bands or exposures, which would make them quite
difficult to identify within the data from the survey alone.  Indeed, in essentially every
survey that has been used for weak lensing analysis to date, the preliminary results indicated
some unforeseen systematics. Despite care taken in the design and planned operations of new 
surveys, we cannot reliably exclude the possibility of unforeseen systematics.

In the case of these three surveys, the systematic errors from each survey are expected to be
very different.  Thus, cross-correlating the shear estimates from one survey with those of
another should remove nearly all shear systematic error contributions, leaving only the true weak
lensing signature.  If this cross-correlation analysis reveals differences from the analysis
within individual surveys, it could signal a problem that needs to be investigated and
mitigated.  Furthermore, the surveys have different redshift distributions for the lensed
galaxies, so a joint cosmological analysis using data from two or more of them will be
complementary.  The joint analysis would also be better able to constrain nuisance parameters
like intrinsic alignments and photometric redshift errors.


As described below, the principal way in which Euclid and WFIRST would benefit from sharing information with LSST is
in the photometric redshift catalog.   Without this additional
data, Euclid and WFIRST will not be able to complete their goals with weak lensing tomography.
LSST will also gain  from the IR band photometry of  WFIRST and Euclid
in terms of photometric redshift determination.

The easiest form of data sharing between surveys is if only catalog information is shared.
For some of the gains mentioned above, this level of sharing is completely sufficient:
cross-correlations and shear calibration, for example.  The deblending improvements would
require an iterative reprocessing of the data, using a space catalog as a prior and
then reprocessing the LSST data with that extra information.  A similar iterative
reprocessing would be necessary for Euclid to properly account for its color-dependent
PSF using the LSST SED estimates as a prior.  For the photometric redshift improvement,
one could just use the multiple catalogs, but to do it properly, it is important to
have access to the pixel data from both experiments at once to properly estimate
robust colors for each galaxy.  It will be a significant technical challenge to
develop a framework that will work with the different kinds of image data in a coherent
analysis.

\subsection{Large-scale structure}

\begin{figure}[!htb]
\begin{center}
\includegraphics[height=3.0in]{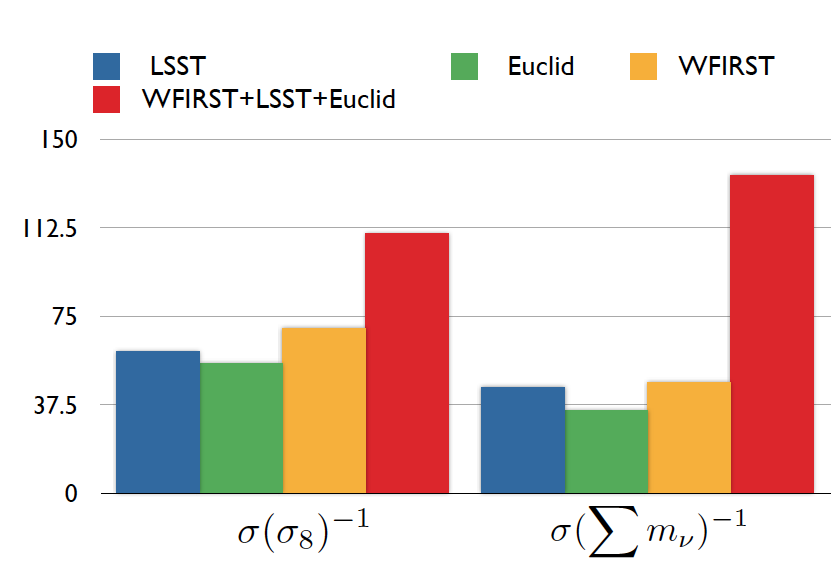}
\caption{\label{fig:comb}
\scriptsize{ The chart shows how the complementarity of LSST, Euclid and WFIRST contributes to  significant improvement in constraints on cosmological parameters.  As described in the text, the improved constraints on 
$\sigma_8$ come from the mitigation of intrinsic alignment and other systematics in weak lensing; the improved constraints on the sum of neutrino masses $\sum m_\nu$ (in eV) comes from the combination of the weak lensing, CMB convergence maps, and galaxy clustering, in particular by reducing the multiplicative bias in shear measurement.
Note that the space based surveys are assumed to have used ground based photometry to obtain photo-z's. 
}}
\end{center}
\end{figure}

Measurements of large-scale structure probe dark energy via baryonic acoustic oscillation features in the galaxy power spectrum. In addition, the full shape and amplitude of the power spectrum measure the clustering of matter, if one can also measure the biasing of galaxies relative to the matter. This provides new avenues to measure the properties of dark energy and probe gravity. 

The upcoming generation of spectroscopic surveys will make detailed maps of the large-scale structure of the universe.
DESI, PFS, Euclid and WFIRST will  focus on measuring galaxy clustering by obtaining spectra of tens of millions of galaxies over redshift ranges from $z\sim 0.4-3.5$. The surveys complement each other using a variety of spectral lines (H$\alpha$, OII, CII) and galaxy types (ELG and LRGs) as tracers and through having distinct, but overlapping, redshift ranges.  DESI and Euclid will be wider surveys, while PFS and WFIRST will go deeper on smaller areas of the sky.

The combination of spectroscopic and lensing data enables a test of general relativity on cosmological scales.
The spectroscopic data provides information about galaxy positions and motions determined  inertial masses moving in the local gravitational potential. The imaging data, by contrast, shows the effect of space-time curvature on the trajectories of photons.
If general relativity is the correct description of gravity then these will agree.
 Spectroscopic and photometric galaxy cluster surveys from WFIRST, LSST, Euclid and DESI, and others, and their cross-correlation with contemporary CMB polarization (CMB lensing) and temperature (CMB lensing and kinetic Sunyaev-Zel'dovich) data will provide coincident, complementary dynamical and weak lensing tests of gravity in halos two to three orders of magnitude more massive than the galaxies within them. Thus we expect powerful tests of gravity by a joint analysis of the three surveys.


We study the complementarity of WFIRST, LSST and Euclid by forecasting constraints on matter clustering (we pick $\sigma_8$ for convenience, which is sensitive to dark energy and is partially degenerate with the parameters, $n_s$, $\Omega_m$), and the sum of neutrino masses in the Universe.  While we concentrate on these  particular parameters, there are many other possibilities considered by the community. 

 Figure~\ref{fig:comb} shows constraints on $\sigma_8$ and the sum of neutrino masses $\Sigma m_\nu$ from the three surveys individually and in combination. We  include a CMB prior from Planck, and marginalize over  cosmological (and galaxy bias) parameters when appropriate. For the space surveys we assume that photo-z's will be obtained with the aid of ground based data. 
WFIRST's densely sampled spectroscopic galaxies will provide a detailed characterization of filamentary structure that is complementary to the lensing measurements provided by imaging surveys.  
  This allows us to understand one of the most elusive systematics in weak lensing surveys -- the intrinsic alignment (IA) of galaxies.
 One can test and validate IA models and marginalize over the free parameters in specific models when we can have a 3D  map of the universe via galaxy spectra.
  In LSST, the lensing sources spans a large redshift range between 0 and 3 and  a large luminosity range.
  With WFIRST and Euclid, we will have spectroscopic sources spanning a large part of this redshift range, which enables a very good construction of filament map of the surveyed volume.  WFIRST offers a much higher density of sources while LSST and Euclid will cover a much large area of sky.
  The joint analysis leads to significant  improvement in the constraint on $\sigma_8$, as shown on the left in Figure~\ref{fig:comb}.  

We also consider the constraints on massive neutrinos which suppress the growth of structure below the free-streaming scale. The combination of WFIRST and Euclid's spectroscopic sample, along with CMB lensing convergence, can be used to calibrate the shear multiplicative bias in LSST  sources. We follow Das, Evrard \& Spergel (2013) in methodology and calculate the improvement in constraints first in the multiplicative bias in shear, and propagate this to the constraints on the sum of neutrino masses that relies on the large sky coverage of LSST.  We assume that a Stage III level CMB  convergence field will be available. We plot the improvement in $\Sigma m_\nu$ on the right in Fig~\ref{fig:comb}.

\subsection{Galaxy clusters}
LSST, Euclid and WFIRST  will use galaxy clusters (primarily via gravitational lensing mass measurements)  as part of their cosmological measurements.  However, both for cosmology and the study of galaxy evolution in clusters, a joint analysis of the data can greatly improve the reach of each mission.  Fundamentally, this is because each of the missions observes different aspects of the emission from clusters.  LSST provides optical colors, useful for cluster identification via photometric redshifts, as well as (for group-scale structures) time delays for lensed objects.  Euclid provides shallow NIR photometry and somewhat deeper NIR  spectroscopy, as well as higher resolution optical imaging.  WFIRST will provide much better resolution images and deep NIR photometry and spectroscopy over a smaller area than Euclid.  Because of the complementarity of these measurements, combining the data has implications for cluster finding, redshift determination, mass measurement and calibration, and the study of galaxy evolution in clusters.

{\it Galaxy cluster finding:}
Over the past decade, techniques have been developed that select galaxy groups and clusters very effectively by isolating overdensities simultaneously in projected position and color space (for example redMaPPer).
These techniques have been optimized for optical wavelengths.  LSST will be extremely efficient at finding groups and clusters of galaxies and estimating their redshifts (although eROSITA will also be an efficient cluster finder, the redshift information available from the LSST imaging will allow selection of samples at fixed redshift for study).  Although redMaPPer is effective to $z\sim 1$, extending its redshift range will require infrared imaging, which would be provided by Euclid or WFIRST.  Given that the number density of detected (as opposed to resolved) sources in LSST and WFIRST are going to be similar, and given the importance of consistent photometry measurements across bands, simultaneous measurements of the pixel data would yield more accurate selection of clusters.

{\it Redshift Determination:}
As discussed below, the photo-z's of galaxies will be greatly improved by combining the optical and infrared imaging of the three missions.  Thus, joint analysis of the LSST+WFIRST/Euclid data gives more accurate cluster sample selection, and allows accurate determination of the background galaxy redshift distribution, which is the largest single source of systematic uncertainty in determination of the mass function from weak lensing mass determinations.

{\it Strong and Weak Lensing:}
Measuring the weak lensing shear requires high resolution imaging.   WFIRST will be able to resolve many more galaxies than LSST, and as discussed above will aid the shear calibration of LSST.  In addition, the conversion of shear to mass depends on redshift information.  In this way, the combination of LSST, Euclid, and WFIRST data will provide much more accurate mass measurements than either will alone.  

Cluster arc tomography is greatly improved by the joint pixel-level analysis of LSST, Euclid and WFIRST data.  For instance, WFIRST will have the angular resolution to discover $\sim 2000$ strong lensing clusters in its survey footprint.  However, most of the arcs will be too faint (and too low surface brightness) to have spectroscopic redshifts.  By contrast, LSST will reach surface brightness limits ($>28.7$ Mag per square arcsecond) sufficient to detect the arcs and obtain photometric redshift estimates, but will not have the resolution to separate the arcs cleanly from the cluster galaxies in many cases.  Thus, the measurement of the tomographic signal from multiple arc systems, which is a strong test of cosmology, will be much stronger from the joint analysis than from Euclid, WFIRST or LSST alone.  Once again, the added value of the pixel data is great, because cluster cores are very dense environments and obtaining valid photometric redshifts will require use of the space-based imaging to disentangle the light contribution form the arcs and cluster galaxies.  

{\it Systematics:}
One of the sources of systematic uncertainty in weak lensing shear measurements from galaxy clusters comes from the blending of sources both with other sources and with cluster galaxies.  The source-source blending is in many ways similar to that encountered in large-scale weak lensing measurements, but in clusters, there is an additional source of bias from blending with cluster galaxies.  This is both radially dependent and (because cluster concentration evolves with redshift) redshift dependent, and directly affects the normalization of the cluster mass function.  Using the WFIRST data to directly measure the deblending bias in the clusters in common between the two surveys would make it possible to correct for the bias.


\subsection{Supernovae}
Each individual Type Ia supernova (SN~Ia) is potentially a powerful  probe of relative distance in the Universe.
Thousands of them provide detailed information about the expansion history of the Universe, in particular allowing us to probe the accelerated expansion and the nature of dark energy.
Hundreds of thousands of them will allow us to probe the nature of dark energy in different regions and environments.
A set spanning the past 10 billion years of cosmic history will allow for a rich and powerful exploration of the nature of the expansion and more recent acceleration of the Universe and thus reveal key insights into the key constituents.

Ground-based SN~Ia searches have provided almost one thousand well-observed SNe~Ia in 20 years of work.  Increasingly powerful ground facilities will rapidly increase this number to thousands per year (e.g., Pan-STARRS, PTF and DES) and then hundreds of thousands per year in the era of LSST.  For a smaller subset (hundreds) of these supernovae, even calibrated spectrophotometric time-series have been obtained (at the nearby SN Factory).  However, the rest-frame emission of SNe~Ia is line-blanketed at UV wavelengths.  Thus, by $z\sim 1$ their light has already redshifted out of the optical atmospheric window easily accessible from the ground.  In addition, because the relative distances across redshifts are  estimated based on relative flux differences across wavelength, accurate cross-wavelength calibration is critical to use supernovae to determine the properties of dark energy.

The stability of WFIRST in space yields significant benefits in both wavelength coverage and flux calibration for thousands of SNe~Ia/year from $0.1<z<1.7$.  The accessibility of the ground allows for the large aperture and high-etendue of the LSST system that can observe 10,000s of SNe~Ia per year from $0.03<z<1$.

Measuring distances from $z\sim0.03$ to $z\sim1.7$ will connect the era when dark energy first made its presence known, through the transition to an accelerating expansion, and up to the present day.  These are the key redshifts over which we will discover hints to the underlying nature of dark energy, and the combination of the high-redshift reach and calibration of WFIRST and the huge numbers and volume coverage of LSST will yield a strong and powerful pairing for SN Cosmology.

The key question for this work will be the control of systematic uncertainties as we explore  a large span of time and changing populations of host-galaxy environments in which the supernovae are born.   The combination of WFIRST, LSST, and other ground-based follow-up instruments will make possible unprecedented rich characterization of each supernova so that low- and high-redshift supernovae can be matched at a more detailed level than  with the broad ``Type~Ia'' definition.   For example, a time series of detailed spectrophotometry could be obtained for the same rest-wavelengths for every supernova.

Modern SN distance analyses are already joint analyses.  The current Hubble Diagrams are constructed from supernovae from different surveys (e.g., Betoule et al. 2014  combined the Supernova Legacy Survey (SNLS), SDSS Supernova Survey, and the current heterogeneous sample of nearby $z<0.05$ SNeIa).
But a true joint analysis should begin at the pixel level and take advantage of all the internal details about calibration and photometry  within each survey.  Specifically, when combining SNeIa from different data sources, a joint analysis is beneficial because of: 1. Consistent pixel-level analysis, in particular photometric extraction; and, 2.  Deep cross-checks of photometric calibration not possible with just extracted SN light curves.

Supernova surveys have distinct science requirements both on supernova discovery and early classification, and on the quality of distance modulus and redshift determinations of discovered supernovae.  Each science requirement propagates into hardware and survey requirements.  Technical requirements for supernova discovery (e.g., large field of view, detection of new point sources) can have a more cost-effective ground-based solution, whereas requirements for photometry and spectroscopy (e.g., high signal-to-noise SN spectral features, infrared wavelength coverage) are only fulfilled with a space observatory.  A case in point is the baseline WFIRST plan where an imaging survey is conducted to identify supernovae for targeted spectroscopic observations.  The large field of view of LSST and the availability of SN photons detectable by LSST out to $z=1.2$ should make the LSST the better instrument for SN discovery, especially at low redshifts where the angular density of potential host galaxies is low.  Assuming the triggering logistics work, this might free up WFIRST to do what it is better at: obtaining supernova spectrophotometric time series.

In addition, there is a benefit of observing the same SN with multiple overlapping surveys.  Such observations provide:
\begin{itemize}
\item Direct cross-instrumental calibration on the source of interest.  (It is important to note, however, that this calibration is not the same as the cross-wavelength calibration that is required when comparing supernovae at very different wavelength.)
\item Expanding observations of the wavelength/temporal range of the supernova.
\item A quantitative assessment of systematic errors.
\end{itemize}

Finally, supernova cosmology analyses require a suite of external additional resources beyond the direct observations of high-redshift supernovae.  Thus, planning for a joint LSST-WFIRST SN cosmology analysis will yield significant gains in combining resources to obtain external data and theory needs.  Some key examples include:
\begin{itemize}
\item Low-z SN set, including possible spectrophotometric time-series follow up.
\item Fundamental flux calibrators (which {\em can} provide the cross-wavelength calibration that is required when comparing supernovae at very different wavelength).
\item High-resolution spectroscopy of host galaxies.
\item Improved SN Ia theory and empirical modeling.
\end{itemize}

\subsection{Strong lensing}

The LSST, Euclid and WFIRST datasets will be particularly complementary in the field of strong gravitational lensing. Galaxy-scale strong lenses can be used as probes of dark energy (via time delay distances or multiple source plane cosmography). The time delays between multiple images in a strong gravitational lens system depend, as a result of the lens geometry, on the underlying cosmology -- primarily the Hubble constant but also, in large samples where internal degeneracy breaking is possible, the dark energy parameters.  (Refsdal 1964, Tewes et al 2013; Suyu et al 2014). 
Lens systems with sources at multiple redshifts may also provide interesting and competitive constraints on cosmological parameters, via the ratio of distance ratios in each system.  

Strong lenses also provide unique information about  dark matter, via the
central density profiles of halos on a range of scales, and the mass function
of sub-galactic mass structures that cause measurable perturbations of
well-resolved arcs and Einstein Rings. Both this science case and strong
lensing cosmology will benefit greatly from the large samples of lenses
detectable in these wide field imaging surveys; we sketch these cases out
below in Section 4, and identify how high fidelity space-based  imaging and
spectroscopy will enable new measurements.

\subsection{Photometric redshifts}

Photometry provides an efficient way to estimate the physical properties of galaxies (i.e.\ their redshifts, spectral types, and luminosities) from a small set of observable parameters (e.g.\ magnitudes, colors, sizes, and clustering).  LSST will depend on these photometric redshifts for all of its major probes of dark energy as it is infeasible to obtain spectroscopic redshifts for the bulk of objects in the LSST sample with any reasonable amount of telescope time.  For the { WFIRST} and { Euclid} missions, three of the key applications will rely on photometric redshifts: measures of the mass power spectrum from analyses of the weak lensing of faint galaxies, breaking of the degeneracies between possible redshift solutions for galaxies exhibiting only a single emission line in grism spectroscopy, and the identification of high redshift Type Ia supernovae based on the photometric redshifts of their host galaxies. In addition, photometric redshifts will play a key role in much of the LSST, { WFIRST}, and { Euclid} extragalactic science where we wish to constrain how the demographics of galaxies change over time.

The high efficiency of photometric redshifts comes at a cost.  The primary features that enable photometric redshift estimation are the transitions of the Balmer (3650 \AA) and Lyman (912 \AA) breaks through a series of photometric filters. Uncertainties in isolating the position of the breaks due to the low spectral resolution of the photometry give rise to a scatter between photometric redshift estimates and the true redshifts of galaxies.  Confusion between breaks (due to sparse spectral coverage and incomplete knowledge of the underlying spectral energy distributions) leads to catastrophic photo-z errors, as seen in Figure~\ref{fig:wfirst_photoz}.

Existing studies have demonstrated that while photometric redshift accuracies of $\sigma_z\sim 0.007(1+z)$ are possible for bright objects with many filters, uncertainties of $\sigma_z\sim 0.05(1+z)$ are more typical when restricted to 5-6 bands of deep imaging.  Catastrophic photometric redshift errors (where the difference between the photometric and spectroscopic redshifts, $\Delta z$, exceeds the $3-\sigma$ statistical error) typically occur in well over 1\% of cases.  As the photo-z scatter and catastrophic failure rate increase, information is degraded and dark energy constraints will weaken.  Furthermore, if there are nonnegligible ($>\sim 0.2\%$) systematic offsets in photo-z's or if the $\Delta z$ distribution is mischaracterized, dark energy inference will be biased at levels comparable to or greater than expected random errors.  As a result, careful calibration and validation of the photometric redshifts will be necessary for the { LSST}, { WFIRST}, and { Euclid} surveys.

\subsubsection{The impact of Euclid and WFIRST near-infrared data on LSST photometric redshifts}

\begin{figure}[h]
\begin{center}
\includegraphics[scale=.45]{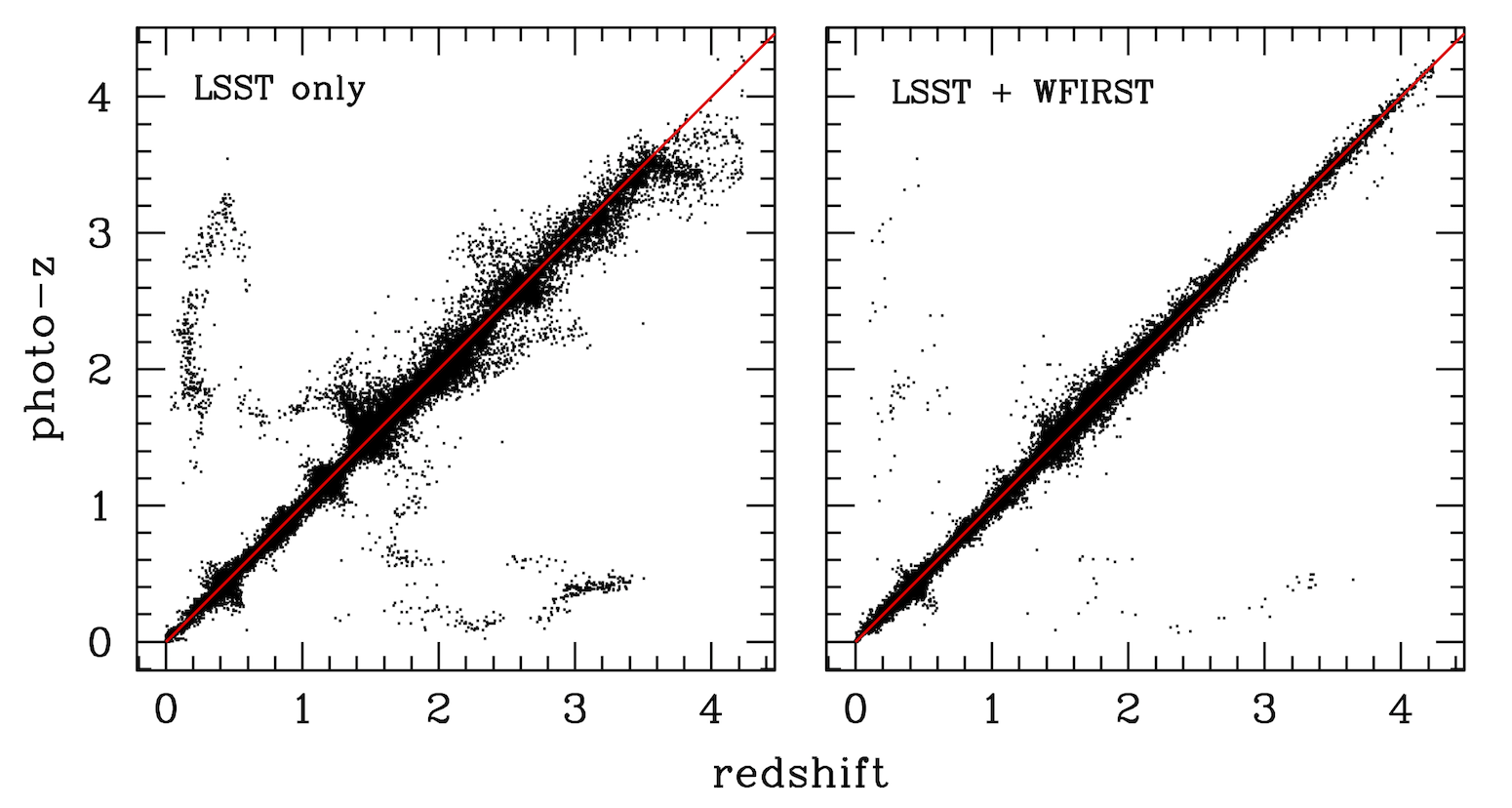}
\end{center}
\caption{\label{fig:wfirst_photoz}\footnotesize A comparison of the relative photometric redshift performance of the LSST optical filters (left panel) with a combination of LSST and WFIRST filters (right panel). The simulated data assumes a 10-year LSST survey and a ``gold sample'' with $i<25.3$. The addition of high signal-to-noise infrared data from WFIRST reduces the scatter in the photometric redshifts by roughly a factor of two (at redshifts $z>1.5$) and the number of catastrophic outliers by a factor of three. These simulations do not account for deblending errors or photometric calibration uncertainties, and assume complete knowledge of the underlying spectral energy distributions of galaxies as an ensemble. }
\end{figure}

The LSST filter system covers the $u, g, r, i, z $ and $y$ passbands, providing substantial leverage for redshift estimation from $z=0$ to $z>6$. For the LSST ``gold'' sample of galaxies, $i<25.3$,
Figure~\ref{fig:wfirst_photoz} shows how the HLS for WFIRST, with four bands comparable in depth to the 10 year LSST survey, (i.e.\ 5$\sigma$ extended source depths of Y$_{AB}$=25.6, J$_{AB}$=25.7, H$_{AB}$=25.7, and F184$_{AB}$=25.2), will significantly improve on the photometric redshift performance from LSST alone.  For example, at redshifts $z>1.5$, where the Balmer break transitions out of the LSST $y$ band and into the WFIRST and Euclid infrared bands, the inclusion of the WFIRST data results in a reduction in $\sigma_z$ by a factor of more than two ($1.5<z<3$), and a reduction in the fraction of catastrophic outliers to $<$2\% across the full redshift range. Euclid's three-band NIR photometry, while shallower, will have a much greater overlap with LSST and will also provide a quantitative improvement in LSST photo-z's. Combining the LSST, WFIRST and Euclid photometric data effectively will depend, however, on the details of the respective filter systems, their signal-to-noise, our ability to extract unbiased photometric measurements from extended sources (e.g.\ the deblending of sources using the higher spatial resolution of the WFIRST data), and the accuracy of the photometric calibration of the data both across the sky and between the near-infrared and optical passbands. 


\subsubsection{Mitigating systematics with WFIRST and Euclid spectroscopy}

\begin{figure}[h]
\begin{center}
\includegraphics[scale=.45]{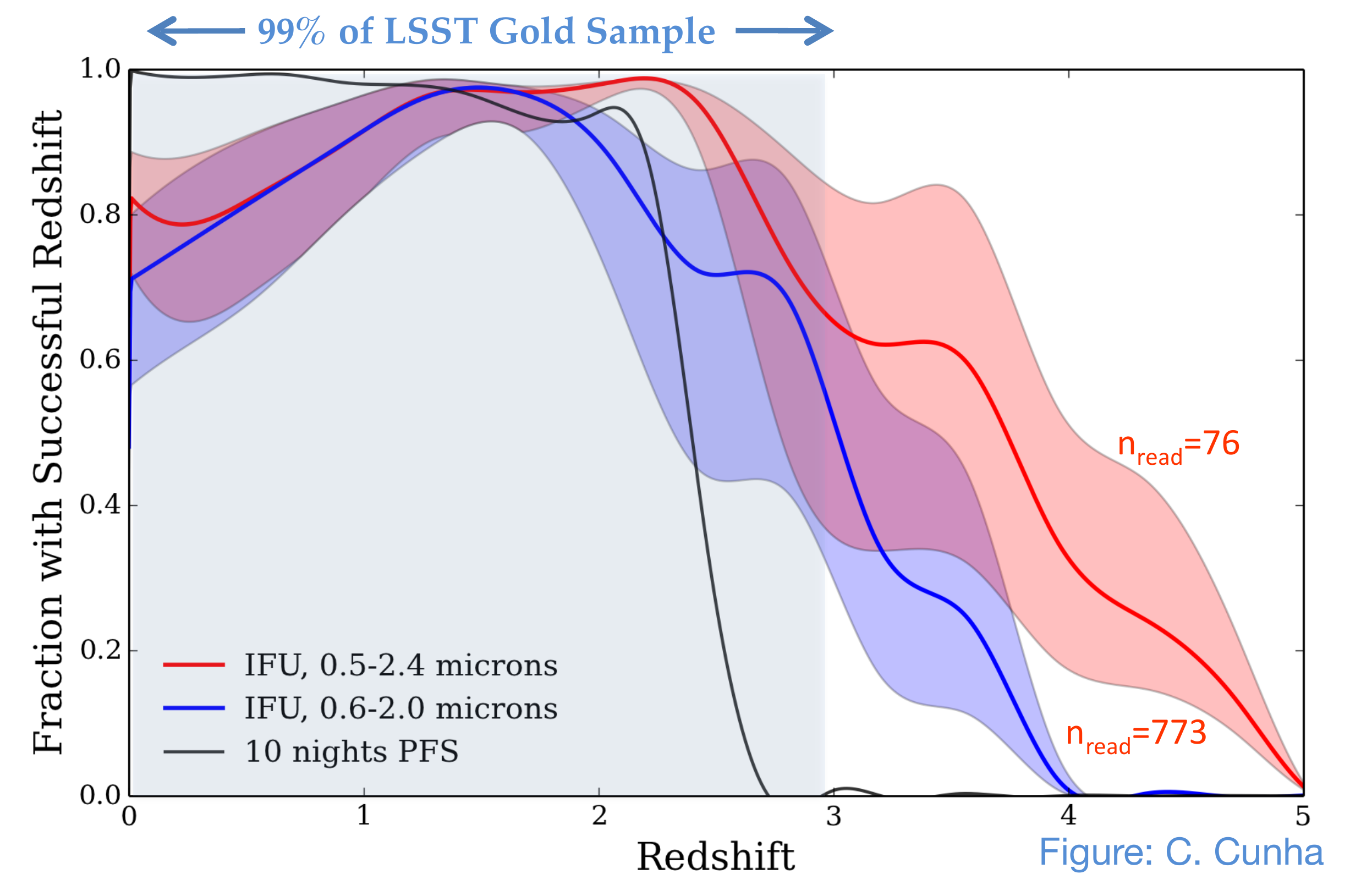}
\end{center}
\caption{\label{fig:ifu}\footnotesize Predictions of the fraction of LSST weak lensing sample objects that would yield a secure (multiple-confirmed-feature) spectroscopic redshift, based either on 1440-second exposure time with {\it WFIRST} (colored regions) or 10 nights' open-shutter-time spectroscopy with the Subaru/PFS spectrograph (black curve)
{\it WFIRST} IFU spectroscopy would provide training redshifts for objects at higher $z$ than are easily accessible from the ground, particularly if read noise per pixel is small (the colored regions indicate a range of feasible scenarios).  Longer exposure times (e.g., in supernova fields or by optimized dithering strategies) could enhance the success rate further.}
\end{figure}

The optimization of photometric redshift algorithms and the calibration of photometric redshift uncertainties both require spectroscopic samples of galaxies. If simple algorithms are used, more than 100 spectroscopic survey regions (of $\sim$0.25 deg$^2$) with at least 300-400 spectroscopic redshifts per region may be required to optimize a photometric redshift algorithm (whether by refining templates and photometric zero points or as input for machine learning algorithms) to ensure that their accuracy is not limited by sample variance in the spectroscopic training set (Cunha et al. 2012); with techniques that take this variance into account, 15-30 fields may be sufficient (Newman et al. 2014).  An ideal training set would span the full range of properties (including redshifts) of the galaxies to which photometric redshifts will be applied.  To the degree to which we do not meet this goal, we can expect that photometric redshift errors will be degraded, weakening constraints on dark energy as well as other extragalactic science.

Current spectroscopic samples fare well; surveys to $R=24.1$ or $i=22.5$ (more than two magnitudes shallower than the LSST ``gold sample'') with 8-10m telescopes have obtained $>$99\% secure redshifts for 21-60\% of all targeted galaxies, and $>$95\% secure redshifts for 42-75\% of the galaxies (incorrect redshift rates above 1\% would lead to photo-z systematics that exceed LSST requirements).
WFIRST spectroscopy can address these limitations in a number of ways.  Depending on the final configuration and dithering strategy for {\it WFIRST}, an object in the LSST weak lensing sample will fall on the IFU field-of-view at least 10\% (and up to $\sim 100\%$) of the time.  Most objects in this sample would yield a successful redshift with a $\sim1440$ sec exposure (see Figure~\ref{fig:ifu}).  For an IFU with a 3`` $\times$ 3`` field-of-view, at least 10,000 spectra down to the LSST weak lensing depth, corresponding to roughly 20,000 down to the {\it WFIRST} limits, would be measured concurrently with the WFIRST HLS (with minimal impact from sample/cosmic variance).  This spectroscopy would have very different incompleteness from ground-based samples, allowing a broader range of galaxies to have well-trained photometric redshifts, with accuracy limited by the imaging depth rather than our knowledge of galaxy SEDs.  

If our spectroscopic samples do not have  a success rate approaching 100\%, accurate characterization of photo-z errors (required for dark energy inference) is likely to instead be based on the cross-correlation between spectroscopic and photometric samples (cf. Newman 2008).  To meet LSST calibration requirements, such an analysis will require secure ($<1\%$ incorrect) spectroscopic redshifts for $\sim 100,000$ galaxies spanning the full redshift range of the photometric samples of interest and an area of at least a few hundred square degrees.  Euclid and WFIRST grism spectroscopy will provide spectra for many millions of galaxies over wide areas (e.g., $2\times10^{7}$ galaxies in the WFIRST HLS). However, this spectroscopy will provide the highly secure, multiple-feature redshifts required for training and calibrating photo-z's in only very limited redshift ranges.  As a result, Euclid and WFIRST grism spectroscopy may contribute to photometric redshift calibration in combination with other datasets spanning the remaining redshift range (e.g., from DESI), but cannot solve this problem on their own.



%

\subsection{Stellar science}

New insights on stellar populations in the Milky Way and nearby galaxies depend critically on gains in photometric sensitivity and resolution.  Thus far, the state of the art imaging surveys of our Galaxy have involved relatively modest tools with respect to today's capabilities.  For example, the SDSS survey is both shallow and operates at low resolution, and IR surveys such as 2MASS and WISE are based on even smaller telescopes operating at even lower resolution.  

The combination of LSST, Euclid, WFIRST and Gaia will transform our view of the Milky Way and nearby galaxies. LSST will map 18,000 square degrees of the night sky every few days, across 6 visible light filters. The imaging depth of LSST will be over 100 times (5 magnitudes) deeper than SDSS, and at higher resolution. WFIRST will overlap some of the same footprint, but will also extend the spectral range of Milky Way surveys to near-infrared wavelengths at even higher spatial resolution (0.1 arcsecond pixels).  Euclid provides much higher spatial resolution than LSST in the visible (aids deblending sources and crowded field photometry) and Gaia addresses the single biggest contribution to the error budget of most stellar population studies aimed at characterizing physical properties; knowledge of the distance of sources.

Much of the stellar population science case for LSST (e.g., from the Science Book) overlaps that from WFIRST (e.g., the community science white papers in the 2013 WFIRST SDT report). These include studies of the initial mass function, star formation histories for Milky Way components, near field cosmology through deep imaging and proper motions of dwarf satellites, discovery and characterization of halo substructure in nearby galaxies, and more. Given the wavelength complementarity alone, most stellar population investigations will be aided by the panchromatic baseline of LSST (and Euclid) + WFIRST (e.g., interpreting SEDs into fundamental stellar properties, increasing the baseline for star formation history fits from multi-band color-magnitude diagrams, confirming memberships of very blue or red stars in either data set, etc.). However, the overall quality of all three data sets for general investigations can be greatly enhanced (and, in a uniform way) by jointly re-processing the pixel data sets for LSST, Euclid, and WFIRST.

To highlight the technical requirements for this joint processing, we point to three LSST and WFIRST "killer apps" in stellar population science,

\begin{itemize}
\item Establishing and characterizing the complete spectrum of Milky Way satellites

\item Testing halo formation models by measuring the shapes, substructure content, masses, and ages of a large set of stellar halos within 50 Mpc

\item Measuring the star formation history, Galactic mass budget, and spatially and chemically-dependent IMF
\end{itemize}

In each of these cases, LSST and Euclid have the advantage with respect to field of view (and cadence). WFIRST has the advantage of higher photometric sensitivity to red giant branch and low mass star tracers, and much higher resolution for star-galaxy separation in the IR.  Joint processing of the three data sets offers the following rewards,

\begin{enumerate}
\item By using the WFIRST astrometry to deblend sources and feed new positions of stars to LSST (and to a lesser extent, Euclid), the LSST and Euclid photometry can be made to go deeper.  In the actual processing, the pixels from LSST, Euclid and WFIRST would be analyzed simultaneously to detect and perform photometry on all detections in any of the more than dozen LSST+Euclid+WFIRST bandpasses.

\item WFIRST photometry alone will be critically insensitive to hot stars (e.g., white dwarfs in the Milky Way, horizontal branch stars in the Local Group).  In most halo and sparse populations, these objects will be easily detected by LSST, and to a shallower depth, Euclid.  Similar to 1.), the LSST photometry can be used to obtain new measurements in the WFIRST data set and to get an infrared color for these sources.

\item As a result of the above, many more sources in the overlapping footprint of the three telescopes will have high-precision panchromatic photometry.  This increased wavelength baseline will yield a much more accurate characterization of the underlying stellar populations (e.g., the sensitivity to age, reddening, and distance variations increases on this baseline).  Note, this is different from matching LSST, Euclid, and WFIRST catalogs since the goal here is to detect sources that otherwise would not have been measured in one of the three data sets.  The biggest gain in our interpretation of the stellar populations to derive physical properties will come from accurate distances to the stellar populations from Gaia.  The different photometric sensitivities is not a concern here, since there are bright tracers in most populations that Gaia can target.

\item The WFIRST morphological criteria of faint sources provides a crucial assessment of LSST's star-galaxy separation.
\end{enumerate}
Many of these joint analyses can be done at the catalog level while others likely require working with the pixels from each of the surveys.

\subsection{Galaxy and quasar science}
The WFIRST photometry will go to a similar depth (on an AB scale) as
the full coadded LSST images. Euclid NIR data, while shallower, will overlap a much larger area of the LSST survey. Thus the combination of LSST, Euclid and WFIRST
will give high S/N 9-band photometry stretching from 3000 \AA  \
to 2
microns (a factor of six in wavelength, twice that of either WFIRST/Euclid or
LSST alone) for the Gold sample ($i<25.3$) of hundreds of millions of galaxies.
This broad wavelength coverage
probes the SEDs of galaxies beyond 4000A to redshifts 3 and beyond,
allowing detailed determination of star formation rates and stellar
masses during the cosmic epoch when galaxies assembled most of their
stars.  (Unobscured) AGN will be identified both from their
SEDs and from their variability in the repeat LSST imaging.

  The Hubble Space Telescope has been able to obtain comparably deep
photometry over a similar wavelength range over tiny areas of sky; for
example, the Cosmic Assembly Near-infrared Deep Extragalactic Legacy
Survey (CANDELS) covers only $0.2$ deg$^2$, while WFIRST will cover an
area ten thousand times larger.  With this much larger sky coverage,
the combination of LSST and WFIRST can explore the evolution of galaxy
properties over this broad range of cosmic history.

  The weak lensing signature from individual galaxies is too small to
be measured, but stacking analyses can be done in fine bins of stellar
mass and redshift, exploring the relationship between dark matter halo
mass, stellar populations, galaxy morphology, AGN activity and
environments during the peak of galaxy assembly.  The exquisite
resolution of the WFIRST images will also allow the role of mergers in
galaxy evolution to be quantified, with galaxy pairs separated by as
little as 3 kpc discernible over essentially the entire redshift range.

  The multi-band photometry will allow both quasars and galaxies at
much higher redshifts to be identified as well.  Current studies with
HST have allowed identification of a handful of galaxy candidates to
redshift 8.  WFIRST + LSST will expand these studies over 10,000 times
the area, looking for objects that are dropouts in the u through y
filters.  The wide area is particularly important for discovering the
rare most luminous objects at these redshifts, that can be followed up
in detail with JWST and the next generation of $\sim$30-meter telescopes.
The study of such objects probes the exit from the Dark Ages, as the
intergalactic medium became reionized.  Understanding the formation of
the first galaxies and the reionization process in detail requires
putting these objects in their large-scale structure context, which is
only possible with samples selected over large solid angles.  In
particular, the expanding bubbles by which reionization is thought to
progress are thought to subtend tens of arcminutes on the sky,
requiring wide-field samples to explore them.

\subsection{The variable universe}
Time domain astronomy encompasses an extremely varied suite of astrophysical topics, from the repeated pulsations of regular variable stars that serve as tracers of Galactic structure and cosmological distances (e.g. RR Lyrae and Cepheids), to transient events that by their nature occur either unpredictably or only once (e.g. stellar flares, supernovae, tidal disruption events). As suggested by these examples, transients and variables occur both locally and at the most distant reaches of our universe, and so comprise a population with both a very high dynamic range in luminosity and a wide span of colors. Time domain astronomy is  a rapidly growing field: the rich public dataset provided by LSST will enable a large part of the astronomy community to analyze  observations of transient and/or variable objects. The potential synergies between LSST, Euclid and WFIRST are as compelling and varied as time domain astronomy itself, but here we highlight some compelling examples.

The combined WFIRST, Euclid and LSST datasets would greatly facilitate the characterization and classification of transient and variable objects. One of the challenges imminent in the coming LSST era is the identification of samples of variable and transient events to be targeted for additional observations. From the transient side, the anticipated high volume of the LSST alert stream requires rapid triage of large numbers of events to identify the most important transients: those events which, if not pursued with followup observations immediately, will be opportunities lost forever. Furthermore, the high volume of events also necessitates the prioritization of the most compelling sources to followup, which in turn requires rapid characterization of an event (where characterization here is distinguished from actual classification into a known category of object). The WFIRST and Euclid datasets will assist with these challenges in a number of ways. First, the deep infrared observations will greatly assist in the identification of the ``foreground fog" of very red Galactic sources: low mass stars and brown dwarfs. While LSST's depth will identify many of these objects in quiescence, some percentage will escape even LSST on account of their very red colors and low luminosities. Therefore, flares on these objects may still create what appear to be cosmological transient events in spite of their local origin. The deep, panchromatic characterization of the static sky will enable the association of transient events with co-located sources, whether those sources are quiescent detections of the same source,  or possible host galaxies of the event whose properties may give clues for a deeper understanding of those transients.

As pointed out in the previous subsection on stellar science, the strength of the combined datasets goes far beyond matching the WFIRST, Euclid and LSST catalogues. Rather it increases the depth of the photometry by making it possible to detect sources that would otherwise not have been detected. The ability to identify (or rule out) infrared counterparts may also be helpful in confirming some of the most elusive transients.  For example, potential electromagnetic counterparts to gravitational wave sources are expected by current models to emit their peak emission in near IR wavelengths. The association of any gravitational wave event detection  with a known galaxy greatly reduces the search volume for EM counterparts to gravitational wave events, and the elimination of faint red background galaxies as putative associated transient kilonova emission will  help confirm potential multi-messenger events.

\newpage
\section{Coordinated observations}

A significant fraction of WFIRST time will be devoted to a Guest Observer program.  In this section of the report, we consider a number
of potential GO programs that will likely follow-up on or extend LSST observations.  While the GO programs will likely be selected by a future Time Allocation Committee (TAC),
we will benefit by preparing for likely observing scenarios.  Here, we consider three examples of joint LSST/WFIRST studies:  WFIRST follow-up of strong lenses and lensed AGNs found in LSST,
joint observations of deep drilling fields, and joint WFIRST/LSST searches for the afterglows of gravitational wave bursts.

\subsection{Measuring dark energy with strong lens time delays}

As described in the previous section, time delays between multiply imaged sources are a powerful cosmological probe. 
LSST will enable time delay distance measurements towards hundreds of lensed AGN and supernovae (Oguri \& Marshall 2010, LSST Science Book). The detection of candidate systems, which have image separations of just 1-3 arcseconds, and confirmation of them by lens modeling, depends critically on the quality and depth of the imaging data, suggesting that there is much to be gained by searching a combined LSST/Euclid/WFIRST imaging dataset. Measuring cosmological parameters with this sample will require high signal to noise ratio and high resolution imaging and spectroscopy of any Einstein Rings discovered in order to constrain the lens mass distributions.   By integrating longer on the most interesting systems, WFIRST can achieve background source densities of over 200 sources/arcmin$^2$, which will enable detailed weak lensing mass maps of these systems.

Joint analysis of all LSST and wide field infrared survey data will improve the completeness and purity of the lens searches in the following ways: 1) a combined catalog of LSST and space-detected objects will produce a more accurate initial list of target lenses, and 2) joint analysis of all the available cutout images will enable better lens modeling, leading to a much cleaner sample of candidates to be confirmed in the followup observations.

Being able to model large proportions of the lens sample precisely with high resolution space-based imaging will enable the surveys themselves to constrain the structure and evolution of massive galaxies across a homogeneously-selected sample, providing important prior information needed by other cosmology probes.

The joint optical and infrared photometry provided by LSST, Euclid and WFIRST will constrain the mass environments of the strong lenses (including the line of sight environment) by providing accurate photometric redshifts and stellar masses for all the relevant galaxies in the lightcone (e.g., Greene at al 2013, Collett et al 2013). 

\subsection{Probing dark matter with grism spectroscopy of lensed AGN}

Measuring the mass function of dark matter subhalos down to small masses (e.g., $10^6$ solar masses)  is a unique test of the CDM paradigm, and of the nature of dark matter in  general. The CDM mass function is expected to be a power law rising as $M^{-1.9}$ down to Earth-like masses, while a generic feature of self-interacting dark matter and warm dark matter models is to introduce a lower mass cutoff. Current limits set that cutoff at somewhere below $10^9 M_\odot$. We know from observations of the environs of the Milky Way that there is a shortage of luminous satellites at those masses, but this may be entirely due to the physics of star formation at the formation epochs of these satellites.  Dark Milky Way satellites at lower masses may be detectable through their dynamical effects on tidal streams, but statistical studies of the one galaxy we live in will always carry high uncertainty.

A powerful way to detect subhalos independent of their stellar content and in external galaxies is via the study of large samples of strong gravitational lenses, in particular the so-called flux ratio anomalies (see Treu 2010, and references therein). Small masses located in projection near the four images of quadruply-lensed quasars cause a strong distortion of the magnification and therefore alter the flux ratios. However, anomalous flux ratios can also be caused by stellar microlensing if the source is small enough. One solution is to observe the narrow line region emission lines, where the lensed quasar emission is sufficiently extended to be unaffected by microlensing. Large samples of thousands of lenses could achieve sensitivity down to $10^6-10^7$ solar masses required to probe the nature of Dark Matter. Unfortunately, at the moment only 3 dozen quads are known, and only a handful of those are bright enough to be observable from the ground (e.g. Chiba et al 2005).

Discovering thousands of quads will require surveys covering thousands of square degrees at sub-arcsecond resolution (Oguri \& Marshall 2010). Between them, Euclid, WFIRST and LSST should discover over 10,000 lensed AGN, one sixth of which will be quad systems. Euclid and WFIRST have the potential to observe of order a thousand of these quads spectroscopically themselves, using narrow near infrared emission lines to avoid the microlensing and enabling the wholesale application of the lens flux ratio experiment  (Nierenberg et al. 2014).

\subsection{Deep drilling fields}

The LSST collaboration plans several ``deep drilling" fields.    While this aspect of the LSST survey is still under design, the
broad plan is to reach $AB \approx 28$ in $ugriz$ bands (and shallower in $y$), which is comparable to the medium depth 
planned for the WFIRST SN survey (J=27.6 and H=28.1 over 8.96 deg$^2$).
The WFIRST deep SN fields will be deeper with  depths of J=29.3 and H=29.4 over 5.04 deg$^2$. LSST
has currently identified four selected fields, each with areas of 9.6 deg$^2$: ELAIS S1, XMM-LSS, COSMOS, and Extended Chandra Deep Field South.
With three days of observation per band, WFIRST can cover each of these deep fields to AB = 28.


The goals of these fields (Gawiser et al., white paper; Ferguson et al., white paper)   are to (1) Test and improve photometric redshifts critical for LSST Main Survey science.
(2) Determine the flux distribution of galaxy populations dimmer than the Main Survey limit that contribute to clustering signals in the Main Survey due to lensing magnification.
(3) Measure clustering for samples of galaxies and Active Galactic Nuclei (AGN) too faint to be detected in the Main Survey
(4) Characterize ultra-faint supernova host galaxies.
(5) Characterize variability-selected AGN host galaxies.
(6) Identify of nearby isolated low-redshift dwarf galaxies via surface-brightness fluctuations.
(7) Characterize of low-surface-brightness extended features around both nearby and distant galaxies.
(8)Provide deep ``training sets", for characterizing completeness and bias of various types of galaxy measurements in the wide survey (e.g. photometry, morphology, stellar populations, photometric redshifts).
All of these goals will be clearly enhanced by complementary data in the infrared.  As already outlined in the report, infrared data will significantly improve the certainty of photo-z's.
WFIRST high spatial resolution will  be particularly powerful in these deeper fields where confusion is likely to be significant. The deblending, color correction and morphology arguments in sec 3 apply even more strongly at the greater depth reached in these fields.

Indeed, WFIRST provides great synergy by providing a high-resolution detection image that will allow LSST to go past the ground-based confusion limit.  This in turn will motivate LSST to go deeper in $gri$ than currently planned for the 
Deep Drilling Fields; by matching the LSST Main Survey filter time distribution, LSST will reach deeper than AB$\sim$29 in these three filters rather than the current baseline of AB$\sim$28.5, providing a better match to the deepest WFIRST imaging.  

It is worth noting that the VISTA VIDEO deep NIR survey regions are located at the centers of the already-approved LSST DDFs, and these will generate initial spectroscopy and multi-wavelength coverage even before WFIRST launches, making them optimal regions for covering deeper in NIR.  However, if WFIRST is able to cover all (or most) of the 9.6 deg$^2$ LSST field-of-view, LSST can respond by skipping the significant fraction of time spent in $y$ band in the current LSST "near-uniform DDF" approach.  Using WFIRST imaging in lieu of LSST imaging in $y$ band (and possibly $z$ band) also offers a big improvement in both resolution and depth, as LSST only reaches AB=28.0,27.0 in $z$ and $y$ in the "near-uniform DDF" approach spending 60\% of DDF observing time in those filters; significantly deeper  $ugri$ images could be obtained if this time is saved.  Full areal coverage by WFIRST of as many LSST DDFs as possible, in $zyJH$, will motivate LSST to devote its observing time in those fields to the deepest possible $ugri$ imaging, enabling a wide range of science with these "ultra-deep drilling" fields.  

\subsection{Coordinated supernovae observations}

If the WFIRST SN medium and deep surveys target the same fields as one of
the LSST deep drilling fields, then coordinated observations would yield 8 band supernova light curves.  The WFIRST IFU measurements of these supernovae would
yield a powerful training set that could greatly enhance the value of
the large LSST SN sample.  These coordinated observations would not require any increase in observing time by either project but would enhance the return of both sets of observations.

\subsection{Variable universe: follow-up on novel targets}

LSST's cadence and depth will allow it to discover new classes of variable objects and ``golden" transients, examples of
known classes of objects that have nearly ideal properties for particular aspects of astronomical study.
WFIRST's  
resolution, spectral coverage,  and wavelength coverage will make it a premier facility for following up on some of these objects in the near infrared.
Since studies of individual targets do not take advantage of WFIRST's large field of view, large ground-based telescopes with high quality
adaptive optics will also be important tools for follow-up observation.

There are a host of potential targets for follow-ups.  Examples include:
\begin{itemize}
\item ``Macronovae" produced by the coalescence of binary neutron stars (Li \& Paczynski 1998, Kulkarni 2005, Bildsten et al. 2007).  These explosive events likely peak in the near-infrared and would be potential targets for both WFIRST imaging and IFU follow-up.
\item  LSST and WFIRST will be our most  powerful combination of telescopes for detecting afterglows of gravitational mergers (Kasliwal and Nissanke 2014; Gehrels et al. 2014). 
Given the LIGO's large error box for each event, there will be many variable objects found at the LSST/WFIRST depths, a program of joint monitoring (perhaps, combined with the deep field program described above) will enable a better characterization of the rich zoo of transients.
\item Gamma Ray Bursts (Rossi et al. 2007). 
\end{itemize}

\newpage

\section{Data management and analysis challenges}

The sections above make clear that in essentially every area of astronomy and cosmology, joint analysis of the LSST, WFIRST and Euclid data will provide  significant benefits. These will be made possible by
linking the data from the different surveys and providing a common access point for interrogating the data in a user-friendly way with the appropriate tools. This would enable scientists to explore
connections between the data sets and stimulate novel community research
leveraging the combined data.  The publication of detailed, accurate documentation
along with the cross-referencing and associated data releases would be the most effective way
to encourage and support the community in pursuing such research. This section describes the issues
involved in carrying out such a program for the photometry and simulations required to analyze and interpret
the data.

\subsection{Multi-resolution photometry}

There are quite a few challenges to producing accurate, consistent photometry
across different filters with widely different resolutions, as would be the
situation for combining LSST data with either WFIRST or Euclid.

We can assume that we will have a catalogue of the positions of objects detected in
at least one band (including single-band detections in u or H), and that we
know the PSF in each band.  There are issues with respect to the chromatic effects
within a single band as well as spatial variation of the PSF across the field of view,
which are important, but we can neglect them for this discussion.
We also assume that blending issues have been taken into account, so we have
a set of pixels in each band that correspond to a single object.

For point sources, photometry is relatively straightforward. Because all point sources have an
identical profile (again, ignoring for now the spatial variation of the PSF)
we may use any aperture to
measure the relative flux of all point sources in an image (a PSF aperture is
optimal).  We may then estimate the image's zero-point which gives us total
fluxes for all sources.  We may repeat this in all the bands, resulting in a
catalogue of fluxes and colors.

Measuring the fluxes of galaxies is harder.  It is not possible to simply
measure the total flux in an aperture large enough to include the entire galaxy
as the measurements are too noisy, and this is not even a well-defined concept.
On the other hand, a small aperture measures a different
fraction of the flux of different galaxies, so we cannot simply apply an
aperture correction as in the stellar case.  Even if a galaxy has the same
profile in each band these fractions are different if the PSF is
band-dependent; if the profiles vary the difficulties are multiplied.

If we can model galaxies adequately, this alleviates the difficulties of flux measurement.  The
PSF-convolved model flux is a reasonable definition of the galaxy's total flux
in each band. However using
model-based photometry is not a panacea, as if the model must be correct. For
example, choosing a model in one band and applying it in another will not yield
the correct color for a galaxy with a red bulge and a blue disk, a scale
length that's a function of band, spiral arms, HII regions, or any other kind of
substructure with a different color from the overall galaxy light.

If the seeing is (or can be made) the same in all bands we are allowed to choose a
single model.  The resulting fluxes are not the galaxy's total fluxes, as
different components are not necessarily weighted correctly, but these ``consistent fluxes" do
represent the fluxes of a certain sample of the galaxy's stellar population.
However, degrading all bands to a lowest common denominator would destroy
much of the extra information present in the higher-resolution space data.
Determining the optimal way to estimate consistent fluxes while keeping as much
of this high-resolution information as possible is still an open research question.

Given the complexity of the problem of measuring galaxy photometry across a decade in wavelength with varying PSFs,
there is no optimal approach.  Different science questions will require different ways of combining the multi-color data.

\begin{itemize}

\item {\it Photometric Redshifts: }
We may use an object's total or consistent fluxes to find a location in
multi-color space and hence its redshift probability function, $p(z)$.
If $p(z)$ is unimodal and sharp,  the two estimators predict nearly the same photo-z. However, in
general this will not be so and $p(z)$ will be different for the two choices of
flux.

\item{\it Choice of Galaxy Model:}
In the rest-frame optical, a simple Sersic or constrained bulge-disk model is
probably sufficient to return reliable fluxes (such models down-weight the
contribution from localized star-forming regions, changing p(z) at least in
principle).  However, modeling galaxies across the broad redshift range covered by WFIRST, Euclid,
and LSST will likely require more complex
models (including, for example, the uv-bright knots that contribute the bulk of
the flux in the far-uv).  Constraining these models is going to be hard; one
approach would be to demand that all components have a consistent $p(z)$.

\item{\it Photometry of Blended Objects}
If our models indeed accurately describe our galaxies then the deblending
problem is no different from the galaxy photometry problem; instead of fitting
multiple components to a single galaxy we fit multiple components to multiple
galaxies, relaxing but not abandoning the constraints on components' $p(z)$.  In practice
it is not clear how well this will work, and there are intermediate schemes (similar
to the deblender used in SDSS) that fit simplified models to the ensemble of objects
to assign the flux to the child objects and then proceed with photometry.  This is
an unsolved problem, but given the upcoming data, we will need to must significant progress in the coming years.

\end{itemize}

\subsection{Joint simulations}
WFIRST, Euclid and LSST will generate a remarkable set of observations in the optical and near-infrared wavebands. Their cosmological interpretation, especially in combination, will be very powerful but also very challenging. Many systematic effects will have to be disentangled from fundamental physics to fully exploit the power of the measurements. Simulations of all aspects of the experiment -- at the cosmological scale, of the instruments, and of the data processing and analysis software -- are critical elements of the systematics mitigation program. Joint analyses of multiple surveys require these simulations to have consistent interfaces, to enable the same realizations of cosmological volumes or multi-wavelength simulated galaxies to be passed through instrument-specific simulators, and ultimately to ensure that the correct correlations and joint statistical properties are represented when mock analyses are performed on the simulated output catalogs.

We focus here on the role of coordinated simulations in two areas -- {\em cosmological simulations} and {\em instrument/pipeline simulations}.

\subsubsection{Cosmological simulations}

Cosmological simulations play key roles within the interpretation of large scale structure surveys: (i) they provide controlled testbeds for analysis pipelines and tools for survey design and optimization via sophisticated synthetic skies populated with galaxies targeted by the specific survey, (ii) they provide predictions for different fundamental physics effects, including dynamical dark energy, modified gravity, neutrinos, and non-trivial dark matter models, (iii) they provide modeling approaches for astrophysical systematics, including gas dynamics, star formation, and feedback effects, and (iv) they provide important information about the error estimates via covariance studies. We will briefly elaborate on the synergies between the simulation programs required for WFIRST and LSST in these four areas in the following.

{\em Synthetic sky maps} --- In order to build synthetic sky maps for Euclid, WFIRST and LSST, large-volume $N$-body simulation with very high force and mass resolution are essential. Next, the simulations have to be populated with galaxies using sub-halo abundance matching or semi-analytic modeling techniques, tuned to match the observed population of galaxies. While the wavelength ranges of Euclid, WFIRST and LSST will be different, a common code infrastructure will be beneficial in some areas (e.g., in exploring model parameter space) and critical in others (e.g., simulating realistic merged multi wavelength catalogs).

{\em Precision predictions for fundamental cosmological statistics} --- The precision requirements for cosmological observables and the fundamental physics to be explored is very similar for Euclid, WFIRST and LSST and a joint program in this area would be very fruitful. Both surveys are promising observations of large scale structure measurements at the sub-percent level, a tremendous challenge for cosmological simulations. Besides the precision challenge, different fundamental physics effects have to be explored systematically for both surveys.

{\em Astrophysical systematics} --- Some of the systematics that have to be accounted for to interpret the observations from Euclid, WFIRST and LSST are common, such as the effects of baryonic physics on the weak lensing shear power spectrum. Since the accuracy requirements are similar, a joint program would be beneficial to both surveys. However, there are significant differences because of the differing ground and space-based nature of the two programs, different instrumentation and associated wavebands, and different galaxy populations and densities. Nevertheless, the underlying sky catalogs can be generated from the same set of simulations.

{\em Covariance matrices} --- Since the Euclid, WFIRST and LSST surveys will have overlapping volumes, combining cosmological constraints from their data sets will require understanding the full covariance matrix of all observables extracted from the data sets, using simulations of the entire relevant volume. While the accuracy requirements for these simulations are much less stringent than for other applications, the large number of simulations required will represent a computational challenge.

\subsubsection{Instrument and pipeline simulations}

Instrument and pipeline simulations are also a critical element of any precision large scale structure program: they inform the relationship between the astronomical ``scene'' in the Universe and the catalogs, parameters, and other information that is recorded in the processed data. They also provide an opportunity to exercise the data reduction and analysis tools prior to first light. The instrument and pipeline tools must be considered together, since in a precision experiment the pipeline is an integral part of the measurement process. Unlike cosmological simulations, these tools are highly instrument-specific: for example, the atmosphere is only included for ground-based observations, the patterns of ghosts and diffraction features are completely different for the LSST, Euclid and WFIRST configurations, and internal optical effects, cross-talk, and readout artifacts are fundamentally different in silicon CCDs versus  NIR arrays.

Despite the differences in instrument simulators, it is critical that they be able to take compatible input data. Precision measurements of galaxy clustering and weak lensing depend on not just counting objects and measuring shapes, but understanding selection biases, blending effects, and error distributions of measured parameters. The combination of WFIRST, Euclid and LSST data will be used to measure photometric redshifts, and for both object-by-object and statistical comparisons of number density and shear. Simulation of these applications requires that the same objects be fed through both pipelines, and hence that the input scene (simulated stars and galaxies) be consistent and have the proper correlations of object properties across wavebands. While proprietary pipelines do not need to be shared outside of the individual constortia, the inputs of these pipelines must remain compatible.

\newpage 

\section{Conclusion}

The scientific opportunity offered by the combination of data from LSST, WFIRST and Euclid
goes well beyond the science enabled by any one of the data sets alone.  The range in wavelength, angular resolution and redshift coverage that these missions jointly span is remarkable. With  major investments
in LSST and WFIRST, and partnership with ESA in Euclid, the US has an outstanding scientific opportunity
to carry out a combined analysis of these data sets.  It is imperative for us
to seize it and, together with our European colleagues, prepare for the defining cosmological pursuit of the 21st century.

The main argument for conducting a single, high-quality reference co-analysis exercise
and carefully documenting the results is the complexity and
subtlety of systematics that define this co-analysis.  Falling back on
many small efforts by different teams in selected fields and for narrow goals
will be inefficient, leading to significant duplication of effort.   

For much of the science, we will need to combine the photometry across multiple wavelengths with varying spectral and spatial resolution -- a technical challenge.
As described in Section 3, the joint analysis can be carried out in ways that have different computational demands.
The most technically demanding joint analysis is to work with pixel level data of the entire area of overlap between the surveys. Many of the goals of a joint analysis require such a pixel-level analysis. If pixel-level joint analysis is not feasible, catalog-level analysis can still be beneficial, say to obtain calibrations of the lensing shear  or the redshift distribution of galaxies. Hybrid efforts are also potentially useful, for example using catalog level information from space for deblending LSST galaxies, or using only a mutually agreed subset of the data for calibration purposes. However  the full benefits of jointly analyzing any two of the surveys can be reaped only through pixel-level analysis.

The resources required to achieve
this additional science are outside of what is currently budgeted for LSST by NSF and DOE,
and for WFIRST (or Euclid) by NASA.  Funding for this science would most naturally
emerge from coordination among all agencies involved, and would be closely orchestrated
scientifically and programmatically to optimize science returns.
A possible model  
would be to identify members of the science teams of each project who would
work together on the joint analysis. The analysis team
would ideally be coupled with an experienced science center acting as a focal point
for the implementation, and simultaneously preparing the public release and documentation
for broadest access by the community.

\end{document}